\documentclass[aps,preprint,pra,floatfix]{revtex4}
\usepackage{graphicx}

\begin{document}

\title{Electronic properties of Mn decorated silicene on hexagonal boron nitride}

\author{T. P. Kaloni$^{1}$, S. Gangopadhyay$^{2}$, N. Singh$^{1}$, B. Jones$^{2}$, and U. Schwingenschl\"ogl$^{1,}$}
\email{udo.schwingenschlogl@kaust.edu.sa,+966(0)544700080}
\affiliation{$^{1}$ PSE Division, KAUST, Thuwal 23955-6900, Kingdom of Saudi Arabia}
\affiliation{$^{2}$ IBM Almaden Research Center, San Jose, California 95120-6099, USA}

\begin{abstract}
We study silicene on hexagonal boron nitride, using first principles calculations. 
Since hexagonal boron nitride is semiconducting, the interaction with silicene is weaker
than for metallic substrates. It therefore is possible to open a 50 meV band gap in the
silicene. We further address the effect of Mn decoration by determining the onsite
Hubbard interaction parameter, which turns out to differ significantly for decoration
at the top and hollow sites. The induced magnetism in the system is analyzed in detail.
\end{abstract}

\maketitle

Silicon based nanostructures, such as two-dimensional silicene (an analogue to graphene) and silicene
nanoribbons are currently attracting the interest of many researchers due to materials
properties that are similar to but richer than those of graphene \cite{verri,olle}. Moreover,
they are advantageous to carbon based nanostructures, as they can be expected to be
compatible with the existing semiconductor industry. It is observed that the electronic
band structure of silicene shows a linear dispersion around the Dirac point, like graphene,
and hence is a candidate for applications in nanotechnology. Due to an enhanced spin orbit 
coupling a band gap 1.55 meV is opened \cite{yao}. A mixture of $sp^2$ and $sp^3$-type
bonding results in a buckled structure, which leads to an electrically tunable
band gap \cite{falko,Ni}. First-principles geometry optimization and phonon calculations
as well as temperature dependent molecular dynamics simulations predict a stable
low-buckled structure \cite{ciraci}. Moreover, stability of silicene under biaxial tensile
strain has been predicted up to 17\% strain \cite{kaloni-jap}.

Silicene and its derivatives experimentally have been grown on Ag and ZrB$_2$ 
substrates \cite{padova,vogt,ozaki}, though there is still discussion about the quality
of the results \cite{Lin}. On a ZrB$_2$ thin film an asymmetric buckling
due to the interaction with the substrate is found, which opens a band gap. In general,
transition metal decorated graphene has been studied extensively both in experiment and theory. 
It has been predicted that 5$d$ transition metal atoms show unique properties with topological 
transport effects. The large spin orbit coupling of 5$d$ transition metal atoms together
with substantial magnetic moments leads to a quantum anomalous Hall effect \cite{blugel}.
A model study also has predicted the quantum anomalous Hall effect for transition metal
decorated silicene nanoribbons \cite{ezawa}. Energy arguments indicate that 
transition metal atoms bond to silicene much stronger than to graphene. As a result,
a layer by layer growth of transition metals could be possible on silicene \cite{Ni1}

The deposition of isolated transition metal atoms on layers of hexagonal boron nitride on 
a Rh(111) substrate has been studied in Ref.\ \cite{fabian}. The authors have
demonstrated a reversible switching between two states with controlled pinning and unpinning of 
the hexagonal boron nitride from the metal substrate. In the first state the interaction
of the hexagonal boron nitride is reduced, which leads to a highly symmetric ring in
scanning tunneling microscopy images, while the second state is imaged as a conventional
adatom and corresponds to normal interaction. Motivated by this work, we present in the 
following a first-principles study of the transition metal decoration of silicene
on hexagonal boron nitride. We will first address the interaction with the substrate
and then will deal with the electronic and magnetic properties of the Mn decorated system,

All calculations have been carried out using density functional theory in the generalized
gradient approximation. We employ the Quantum-ESPRESSO package \cite{paolo}, taking into
account the van-der-Waals interaction \cite{grime}. The calculations are performed with 
a plane wave cutoff energy of 816 eV, where a Monkhorst-Pack $8\times8\times1$ k-mesh is
used to optimize the crystal structure and to obtain the self-consistent electronic
structure. The atomic positions are relaxed until an energy convergence of 10$^{-7}$ eV
and a force convergence of 0.001 eV/\AA\ are reached. To study the interaction of the
silicene with the substrate, we employ a supercell consisting of a $2\times2$ supercell
of silicene on top of a $3\times3$ supercell of hexagonal boron nitride. We have tested
the convergence of the results with respect to the thickness of the substrate by taking
into account 2, 3, 4, and 6 atomic layers of $h$-BN, finding only minor differences (in particular
concerning the splitting and position of the Dirac cone) because of the
inert nature of the substrate. A thin substrate consequently turns out to be fully
sufficient in the calculations. Moreover, the $2\times2$ supercell of silicene fits well on
the $3\times3$ supercell of the substrate with a lattice mismatch of only 2.8\%. When we consider
Mn decorated silicene we use a larger supercell that contains 16 Si in a layer over 18 B and 18 N. 
While the onsite Hubbard parameter for 3$d$ transition metal atoms is
known to be several eV, we explicitly calculate the value in the present study for the
different adsorption sites in order to obtain accurate results for the electronic and
magnetic properties of the Mn decorated system.

In general, the lattice mismatch of 2.8\% between silicene and hexagonal boron
nitride can be expected to be small enough to avoid experimental problems with a controlled
growth. Moreover, accurate measurements of materials properties can be 
difficult to achieve on metallic substrates, whereas the interaction is reduced on
semiconducting substrates. Our calculated binding energy for the interface between silicene
and hexagonal boron nitride is only 100 meV per Si atom, as compared to typically 500
meV per Si atom for an interface to a metallic substrate. Experimental realizations
of graphene based electronic devices using hexagonal boron nitride as substrate on a Si
wafer support are subject to various limitations, such as a poor on/off ratio \cite{kim}.
However, on this substrate graphene exhibits the highest mobility \cite{dean} and a
sizable band gap \cite{Gweon,Ruge,jmc}. Since silicene resembles the structure of
graphene, synthesis on hexagonal boron nitride thus has great potential.

The structural arrangement of the system under study is depicted in
Fig.\ 1(b), together with the charge redistribution introduced by the interaction with
the substrate. We obtain Si$-$Si bond lengths of 2.24 \AA\ to 2.26 \AA\ and a
buckling of 0.48\AA\ to 0.54 \AA, which is slightly higher than in free-standing silicene
\cite{yao,ciraci}. For the angle between the Si$-$Si bonds and the normal
of the silicene sheet we observe values of 113$^{\circ}$ to 115$^{\circ}$, again
close to the findings for free-standing silicene (116$^{\circ}$). The optimized distance
between the silicene and hexagonal boron nitride sheets forming the interface turns out
to be 3.57 \AA, which is similar to the distance at the contact between graphene and
hexagonal boron nitride. In addition, the interlayer distance within the hexagonal
boron nitride amounts to be 3.40 \AA, whereas in a bilayer configuration values of 3.30 \AA\
to 3.33 \AA\ have been reported \cite{Marini,shi}. 

The interaction between silicene and hexagonal boron nitride recently has been addressed
by Liu and coworkers \cite{Zhao}, who have reported a perturbation of the Dirac cone
with an energy gap of 4 meV. This study has taken into account only a single layer of
hexagonal boron nitride as substrate, so that a more realistic description may yield
a different result. Indeed, we observe a perturbed Dirac cone with an energy gap of
50 meV in the band structure shown in Fig.\ 1(a). The $\pi$ and $\pi^*$ bands forming
the Dirac cone are due to the $p_z$ orbitals of the Si atoms, while the bands related to
the B and N atoms are located about 0.5 eV above and 1 eV below the Fermi energy. We find
a small but finite charge redistribution across the interface to the substrate; see the
charge density difference isosurfaces plotted in Fig.\ 1(b). As a result the Dirac cone
is perturbed and the 50 meV energy gap is realized, which can be interesting for
nanoelectronic device applications, in particular because an external electric field
can be used to tune the gap. The isosurface plot also demonstrates that the Si atom
closest to a B atom is subject to the strongest charge transfer, while for all other Si
atoms charge transfer effects are subordinate due to longer interatomic distances.

\begin{figure*}[t]
\includegraphics[width=0.5\textwidth]{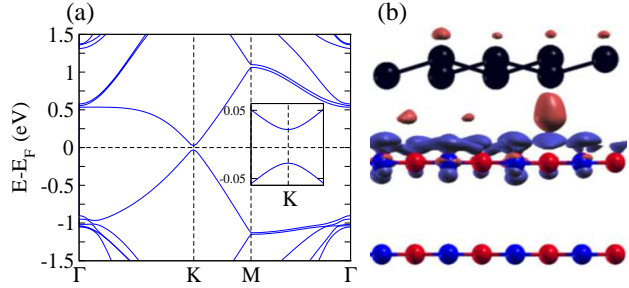}
\caption{(a) Electronic band structure and (b) charge transfer for silicene on a bilayer of BN (side view). 
The isosurfaces correspond to isovalues of $\pm5\times10^{-4}$ electrons/\AA$^3$. 
The black, blue, and red spheres denote Si, N, and B atoms, respectively. Red and blue isosurfaces refer to positive 
and negative charge transfer. Note the significant charge transfer of the Si closest to the BN.}
\end{figure*}

The possible decoration sites for a Mn atom on silicene can be classified as top, bridge,
and hollow. Decoration at the bridge site is not considered in the following because
the Mn atom immediately transfers to the top site. A side view of the relaxed structure
for Mn decoration at the top site is given in Fig.\ 2(a), together with a spin density map. 
We obtain the onsite interaction parameter using a constraint-GGA method \cite{epl}, 
and calculate the values of 3.8 eV for the Mn atom at the top site and 4.5 eV for the Mn 
atom at the hollow site. For the top site configuration, 
structural optimization reveals that the Mn atom moves close to an original
Si position and thereby strongly displaces this Si atom, resulting in a short Mn$-$Si bond
length of 2.43 \AA. Moreover, the Mn atom is bound to three Si atoms with equal bond lengths
of 2.45 \AA. Si$-$Si bond lengths of 2.24 \AA\ to 2.28 \AA\ are observed, which corresponds
to a slight modification as compared to the pristine configuration. The buckling of the
silicene, on the other hand, is strongly altered, now amounting to 0.45 \AA\ to 0.67 \AA.
Accordingly, angles of 113$^\circ$ to 117$^\circ$ are found between the Si$-$Si bonds
and the normal of the silicene sheet. The height of the Mn atom above the silicene sheet
is 1.30 \AA. Finally, we note that the separation between the atomic layers of the hexagonal
boron nitride is virtually not modified by the Mn decoration. 

A side view of the relaxed structure for Mn decoration on the hollow site is shown
in Fig.\ 2(b). In this case, the Mn atom does not displace a specific Si atom but
stays close to the center of the Si hexagon. It is bound equally to the
neighboring Si atoms with bond lengths of 2.40 \AA\ to the upper three and
2.77 \AA\ to the lower three Si atoms. A buckling of 0.46 \AA\ to 0.62 \AA,
Si$-$Si bond lengths of 2.23 \AA\ to 2.28 \AA, and angles to the normal of
112$^{\circ}$-117$^{\circ}$ are obtained. The Mn atom is located 1.01 \AA\ above the
silicene sheet and the separation between the atomic layers in the hexagonal boron
nitride is slightly increased to 3.44 \AA. In contrast, the distance between silicene
and substrate here amounts to 3.55 \AA\ and thus is significantly larger than in
the case of decoration at the top site, because in the latter case one Si atom is displaced
from the silicene sheet, which modifies the distance to the substrate. The calculated
total energies indicate that decoration at the hollow site is by 33 meV favorable as compared
to decoration at the top site.

We find total magnetic moments of 4.56 $\mu_B$ and 3.50 $\mu_B$ per supercell 
for Mn decoration at the top and hollow sites, a reduction of spin from the free Mn 
value of 5.0 unpaired electrons. The magnetization reduction is notable on the hollow site, 
which can be seen from Fig.\ 2 and 3 to involve greater immersion in and hybridization with 
the Si than the top site. By far the largest contribution to the magnetic moment comes from 
the Mn atom and only small moments are induced on the Si atoms.
This can be clearly seen in the spin density maps presented in
Figs.\ 3(a) and (b). For Mn decoration at the top site we obtain a Mn moment of
4.40 $\mu_B$ and a total of 0.16 $\mu_B$ from all the Si atoms, whereas for
decoration at the hollow site the Mn moment amounts to 4.16 $\mu_B$ and the Si
atoms contribute a total $-0.66$ $\mu_B$. These results indicate that the Mn and Si 
moments are ordered ferromagnetically and antiferromagnetically for decoration
at the top and hollow sites, respectively.

\begin{figure*}[t]
\includegraphics[width=0.25\textwidth]{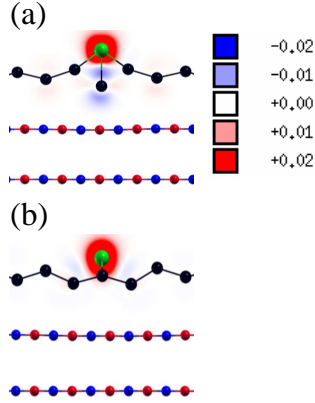}
\caption{The spin density map for silicene decorated by Mn at the
(a) top and (b) hollow site of the $h$-BN substrate. The hollow site is energetically favorable.}
\end{figure*}

In Fig.\ 3 we address the density of states (DOS) for decoration at the (a) top and
(b) hollow sites. The left panel of the figure shows the total DOS and the right panel
the partial DOSs of the Mn $3d$ and $4s$ orbitals. In contrast to pristine silicene
(band gap of 1.55 meV \cite{yao}), the DOSs show a region without states around
0.5 eV below the Fermi energy. This observation corresponds to an $n$-doping due to the
mentioned charge transfer from Mn to silicene. Closer inspection of the partial DOSs for
decoration at the top site shows that the spin majority $s$ and $d_{3r^2-z^2}$ as well
as the spin minority $d_{3r^2-z^2}$, $d_{x^2-y^2}$, and $d_{xy}$ states contribute
in the vicinity of the Fermi energy, while there are essentially no contributions
from the $d_{zx}$ and $d_{zy}$ states. A sharp Mn peak is obsvered about 0.8 eV, which
is due to the spin minority $d_{3r^2-z^2}$ states. For decoration at the hollow site
almost exclusively the spin majority $s$ and spin minority $d_{x^2-y^2}$ and $d_{xy}$ states
contribute around the Fermi energy. Two less pronounced DOS peaks appear 0.75 
eV below the Fermi energy, contributed by the spin minority $d_{xy}$ and $d_{x^2-y^2}$ states.

\begin{figure*}[t]
\includegraphics[width=0.95\textwidth]{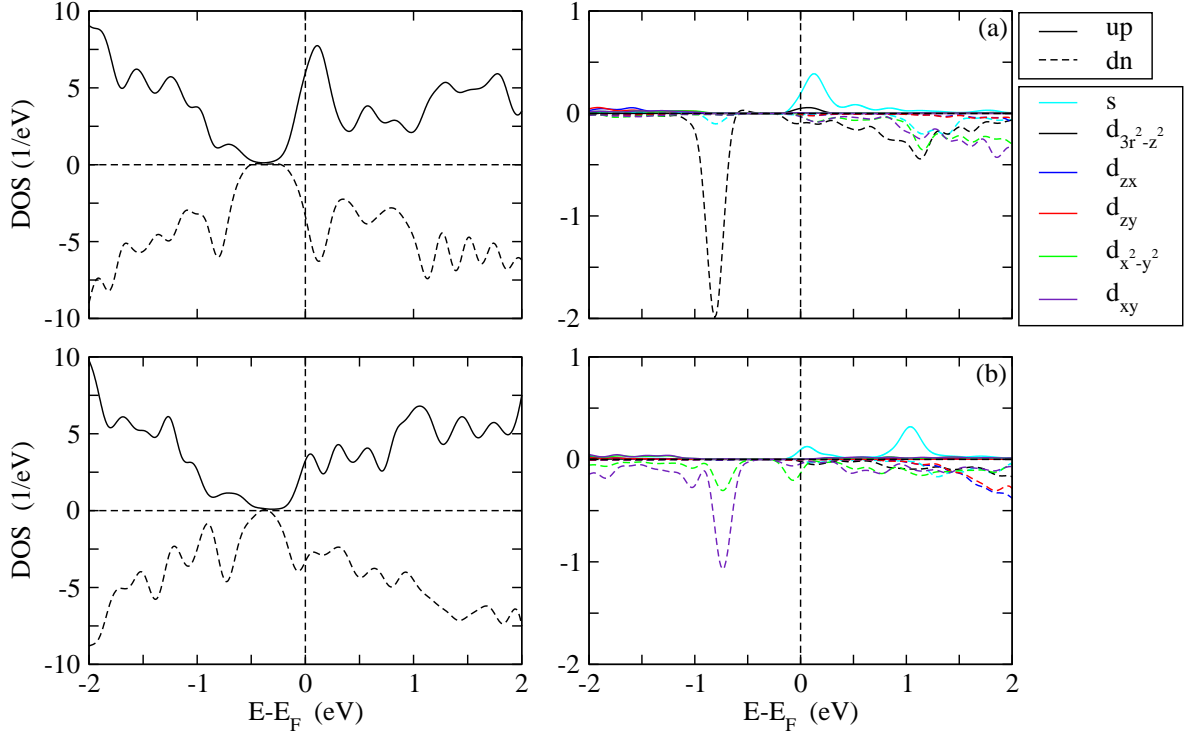}
\caption{Total (left) and Mn partial (right) densities of states of silicene decorated by Mn at
the (a) top and (b) hollow site of the $h$-BN substrate.}
\end{figure*}

In conclusion, we have employed density functional theory to discuss the structure and
chemical bonding of silicene on hexagonal boron nitride. The interaction results in
a band gap of 50 meV. Furthermore, we have calculated the onsite Hubbard interaction
parameter for Mn decoration at the top and hollow sites of the silicene, finding values
of 3.8 eV and 4.5 eV, respectively. The electronic and magnetic properties of Mn decorated
silicene have been studied in detail. In particular, magnetic moments of 3.50 $\mu_B$
and 4.56 $\mu_B$, respectively, have been obtained for Mn decoration at the top and hollow
sites. Interestingly, the orientation between the Mn and induced Si moments is ferromagnetic
in the former and antiferromagnetic in the latter case.


\begin{thebibliography}{50}
\bibitem{verri} G. G. Guzm\'an-Verri and L. C. L. Y. Voon, Phys. Rev. B {\bf 76}, 075131 (2007). 

\bibitem{olle}S. Lebegue and O. Eriksson, Phys. Rev. B {\bf 79}, 115409 (2009). 

\bibitem{yao} C.-C. Liu, W. Feng, and Y. Yao, Phys. Rev. Lett. {\bf 107}, 076802 (2011).

\bibitem{falko}N. D. Drummond, V. Z\'olyomi, and V. I. Fal$'$ko, Phys. Rev B {\bf 85}, 075423 (2012).
 
\bibitem{Ni}Z. Ni, Q. Liu, K. Tang, J. Zheng, J. Zhou, R. Qin, Z. Gao, D. Yu, and J. Lu, Nano Lett. {\bf 12}, 113 (2012).

\bibitem{ciraci}S. Cahangirov, M. Topsakal, E. Akt\"urk, H. Sahin, and S. Ciraci, Phys. Rev. Lett. {\bf 102}, 236804 (2009).

\bibitem{kaloni-jap} T. P. Kaloni, Y. C. Cheng, and U. Schwingenschl\"ogl, J. Appl. Phys. {\bf 113}, 104305 (2013).

\bibitem{padova}P. De Padova, C. Quaresima, C. Ottaviani, P. M. Sheverdyaeva, P. Moras, C. Carbone, D. Topwal, B. Olivieri, A. Kara, H. Oughaddou, B. Aufray, and G. Le Lay, Appl. Phys. Lett. {\bf 96}, 261905 (2010).

\bibitem{vogt}P. Vogt, P. De, C. Quaresima, J. Avila, E. Frantzeskakis, M. C. Asensio, A. Resta, B. Ealet, and G. Le Lay,  Phys. Rev. Lett. {\bf 108}, 155501 (2012).

\bibitem{ozaki}A. Fleurence, R. Friedlein, T. Ozaki, H. Kawai, Y. Wang, and Y. Yamada-Takamura, Phys. Rev. Lett. {\bf 108}, 245501 (2012). 

\bibitem{Lin}C.-L. Lin, R. Arafune, K. Kawahara, M. Kanno, N. Tsukahara, E. Minamitani, Y. Kim, M. Kawai, and N. Takagi, Phys. Rev. Lett. 
{\bf 110}, 076801 (2013).

\bibitem{blugel} H. Zhang, C. Lazo, S. Bl\"ugel, S. Heinze, and Y. Mokrousov, Phys. Rev. Lett. {\bf 108}, 056802 (2012).   

\bibitem{ezawa} M. Ezawa, Phys. Rev. Lett. {\bf 109}, 055502 (2012).

\bibitem{Ni1} X. Lin and J. Ni, Phys. Rev. B {\bf 86}, 075440 (2012).

\bibitem{fabian} F. D. Natterer, F. Patthey, and H. Brune, Phys. Rev. Lett. {\bf 109}, 066101 (2012).                        

\bibitem{paolo}P. Giannozzi, S. Baroni, N. Bonini, M. Calandra, R. Car, C. Cavazzoni, 
D. Ceresoli, G. L. Chiarotti, M. Cococcioni, I. Dabo, A. Dal Corso, S. de Gironcoli, S. Fabris, 
G. Fratesi, R. Gebauer, U. Gerstmann, C. Gougoussis, A. Kokalj, M. Lazzeri, L. Martin-Samos, N. Marzari,
F. Mauri, R. Mazzarello, S. Paolini, A. Pasquarello, L. Paulatto, C. Sbraccia, S. Scandolo, G. Sclauzero, 
A. P. Seitsonen, A. Smogunov, P. Umari, and R. M. Wentzcovitch, J. Phys.: Condens. Matter {\bf 21}, 395502 (2009).

\bibitem{grime}S. Grimme, J. Comput. Chem. {\bf 27}, 1787 (2006).

\bibitem{kim}K. Kim, J.-Y. Choi, T. Kim, S.-H. Cho, and H.-J. Chung, Nature {\bf 479}, 338 (2011).

\bibitem{dean}C. R. Dean, A. F. Young, I. Meric, C. Lee, L. Wang, S. Sorgenfrei, K. Watanabe, T. Taniguchi, P. Kim, K. L. Shepard, and 
J. Hone, Nat. Nanotechnol. {\bf 5}, 722 (2010).

\bibitem{Gweon}S. Y. Zhou, G.-H. Gweon, A. V. Fedorov, P. N. First, W. A. de Heer, D.-H. Lee, F. Guinea, A. H. Castro Neto, 
and A. Lanzara, Nature Mater. {\bf 6}, 770 (2007). 

\bibitem{Ruge}R. Quhe, J. Zheng, G. Luo, Q. Liu, R. Qin, J. Zhou, D. Yu, S. Nagase, W.-N. Mei, Z. Gao, and J. Lu, NPG Asia Materials {\bf 4}, 1 (2012).

\bibitem{jmc} T. P. Kaloni, Y. C. Cheng, and U. Schwingenschl\"ogl, J. Mater. Chem. {\bf 22}, 919 (2012).

\bibitem{Marini} A. Marini, P. Garcia-Gonzalez, and A. Rubio, Phys. Rev. Lett. {\bf 96}, 136404 (2006).

\bibitem{shi} Y. Shi, C. Hamsen, X. Jia, K. K. Kim, A. Reina, M. Hofmann, A. L. Hsu, K. Zhang, H. Li, Z.-Y. Juang, M. S. Dresselhaus, 
L.-J. Li, and J. Kong, Nano Lett. {\bf 10}, 4134 (2010). 

\bibitem{Zhao}H. Liu, J. Gao, and J. Zhao, J. Phys. Chem. C {\bf 117}, 10353 (2013).

\bibitem{epl} G. K. H. Madsen and P. Nov\'ak, EPL {\bf 69}, 777 (2005). 
\end{thebibliography}
\end{document}